# Refuting Strong AI:
# Why Consciousness Cannot Be Algorithmic

Andrew Knight
aknight@alum.mit.edu

## Abstract

While physicalism requires only that a conscious state depends entirely on an underlying physical state, it is often assumed that consciousness is algorithmic and that conscious states can be copied, such as by copying or digitizing the human brain.  In an effort to further elucidate the physical nature of consciousness, I challenge these assumptions and attempt to prove the Single Stream of Consciousness Theorem ("SSCT"): that a conscious entity cannot experience more than one stream of consciousness from a given conscious state.  Assuming only that consciousness is a purely physical phenomenon, it is shown that both Special Relativity and Multiverse theory independently imply SSCT and that the Many Worlds Interpretation of quantum mechanics is inadequate to counter it.  Then, SSCT is shown to be incompatible with Strong Artificial Intelligence, implying that consciousness cannot be created or simulated by a computer.  Finally, SSCT is shown to imply that a conscious state cannot be physically reset to an earlier conscious state nor can it be duplicated by any physical means.  The profound but counterintuitive implications of these conclusions are briefly discussed.

# 1. Introduction

Will computers ever be conscious? Will it ever be possible to cheat death by uploading one's consciousness to a computer? Will it ever be possible to teleport oneself by copying one's brain on a distant planet? Is consciousness algorithmic? Can a conscious state be duplicated? In this paper, I aim to propose and prove a theorem about consciousness and then to discuss some of its several surprising implications.

**Single Stream of Consciousness Theorem ("SSCT"): A conscious entity cannot experience more than one stream of consciousness from a given conscious state.**

To prove SSCT, I need not define "conscious" or "stream of consciousness" nor prove that a conscious entity (i.e., a person) experiences a stream of consciousness at all. Rather, I simply need to show that *if* a conscious entity experiences a conscious state, and *if* it experiences a stream of consciousness, then it cannot experience more than one stream of consciousness from that state. If SSCT is correct, then if I am conscious now, I can expect not to experience more than one stream of consciousness moving forward. While there are lots of different paths that my present conscious state *could* take, I can only have one *actual* future, if I have a future at all. Though SSCT seems intuitively correct[1], I will attempt to prove it in Sections 2 and 3. In Sections 4 and 5 I will discuss two particularly interesting implications of SSCT, including that consciousness cannot be simulated on a computer nor can it be physically reset or duplicated.

In this paper I will assume that consciousness is a purely physical phenomenon ("Physical Consciousness"). Specifically, a person's experience of consciousness depends entirely on an appropriate physical configuration of matter, and the person's experience of a stream of consciousness depends entirely on a physical evolution of an appropriate physical configuration of matter. In other words, whether consciousness supervenes on physical state, or arises or emerges or is caused by physical state, I assume that different conscious experiences could not arise from the same physical configuration of matter.

A related assumption, and the rough equivalent to the notion of Strong Artificial Intelligence (see, e.g., Searle 1980), is that consciousness is algorithmic and depends entirely on a process flow or computation ("Algorithmic Consciousness"). Specifically, a person's experience of consciousness is entirely an emergent property of execution of an appropriate algorithm, the algorithm characterized by a process flow; the person's experience of a stream of consciousness is entirely an emergent property of the process flow.

The assumption that consciousness is purely physical has far more significant implications than is often recognized. For example, if a conscious experience arises from the particular configuration of some isolated physical matter – a brain in a vat, for example – then an identical configuration of physical matter must produce precisely the same conscious experience. Everything about the two experiences necessarily must, by the very assumption of supervenience on the physical, be identical. The two physical configurations must produce exactly the same person having exactly the same subjective conscious experience.[2] If one is experiencing a tingling emotional excitement while skydiving, the other is experiencing a tingling emotional excitement while skydiving. One's perception of identity and self-awareness *is* the other's perception of identity and self-awareness. One's perception of "now" *is* the other's perception of "now" – and this fact is independent of whether or not the two conscious entities happened to be identical at the same *time*.[3] Indeed, if the two conscious entities are spacelike separated, Special Relativity[4] guarantees that there is no fact about simultaneity or temporal order anyway. There is no physical sense in which they are different; they therefore define exactly *one* subjective experience.

---

[1] One interesting but inadequate rebuttal to SSCT is the Many Worlds Interpretation of quantum mechanics, which will be addressed in the Appendix.

It may be rebutted that two otherwise identical physical systems separated by space or time are distinct conscious entities that are not, as viewed from outside the universe, numerically identical. This rebuttal fails for several reasons. First, as currently understood, the laws of physics apply equally throughout space and time; whatever conscious state arises from the physical laws acting on one configuration of matter will be the same as that arising from the other. Second, the rebuttal distracts from the crux of the issue: whether a particular conscious experience of a particular person could occur elsewhere in spacetime. When a scientist wonders whether it will become technologically possible to teleport himself to another planet, or to upload his mind to a computer, or to escape death by simulating his consciousness in a million years, he is not interested in recreating a mind that is *pretty darn similar* to his or even seemingly identical as measured externally. He is interested in recreating *his* mind. He does not care whether *some* conscious person will awaken on a distant planet, he cares whether *he* will awaken on a distant planet. If physicalism is to have meaning, it must be accepted that among two identical isolated physical systems, if one produces a conscious person, the other produces the *same* conscious person. After all, if the only way to produce one's particular identity and conscious experience is to create an entire duplicate universe with precisely the same history as our own, then physicalism has no more explanatory power than an appeal to the Divine.

## 2. Special Relativity Implies SSCT

I will now attempt to prove that SSCT is a direct consequence of Special Relativity. Let us consider a few thought experiments to discover some of the implications of the assumption of

---

[2] To borrow from the philosophy literature on personal identity, the conscious entity arising from each of the configurations must be both "type" and "token" identical, because any difference between them must be due to a physical difference, already posited to be nil.

[3] Arnold Zuboff agrees: "This experience [across brains] of being you, here, *now*, would be numerically the same *whenever*, as well as wherever, it was realized" (1990).

[4] Special Relativity asserts that information cannot travel faster than light. Two events in spacetime are "spacelike" if information from one cannot reach the other without exceeding the speed of light, "lightlike" if such information must travel at the speed of light, and "timelike" otherwise.

Physical Consciousness and its relationship to SSCT. Imagine that your brain (or whatever physical system determines your conscious state) at time $t_1$ is in physical state $S_1$ such that you are conscious – that is, state $S_1$ is an appropriate physical configuration of matter from which emerges your conscious state $C_1$. You have experienced a stream of consciousness due to the physical evolution of matter that led to state $S_1$. Immediately after time $t_1$, your brain is disassembled and soon thereafter, at time $t_2$, reassembled back to state $S_1$, from which it physically evolves to state $S_2$, from which emerges your conscious state $C_2$, at time $t_3$. Note that reassembly at time $t_2$ to state $S_1$ necessarily produces the *same* conscious state $C_1$. It should not make any difference whether the reassembly occurs at the same physical location or some timelike distance away, nor should the time interval between $t_1$ and $t_2$ matter. Nor should it matter whether your brain is destroyed and then recreated from *different* matter, so long as it is the same *configuration* of matter, given that two particles in the same quantum state are physically identical and indistinguishable. Given this hypothetical, what do you experience at time $t_1$?

It seems apparent to me that you would experience a stream of consciousness flowing directly from $C_1$ at $t_1$ to $C_2$ at $t_3$ via $C_1$ at $t_2$, as if the stream skipped over the period between $t_1$ and $t_2$. And it *would* be you, necessarily, because Physical Consciousness guarantees that if state $S_1$ corresponds to a conscious person, then the physically identical state later must correspond to the *same* conscious person in the *same* conscious state. This result would obtain even if $t_2$ were a million years after $t_1$ and the reassembly occurred thousands of light-years away. You would, it seems, experience a continuous stream of consciousness directly from $t_1$ to $t_2$ and through to $t_3$.

Next, imagine that *two* copies of state $S_1$ are created at time $t_2$, one called Configuration A and the other B, and each physically evolves to the same state $S_2$ at time $t_3$ (and in the same manner). Assume that while the locations of creation of both A and B are timelike to your location at $t_1$, they are spacelike to each other. It should be clear that *both* configurations would be you – that is, you would experience consciousness emerging from the (identical) states of both configurations. Of course, this example is trivial because, given that these physical configurations are stipulated to evolve identically, your experience of "two" states is really just one conscious experience. The fact that the configurations are spacelike separated would initially seem to be problematic because there is not enough time for a speed-of-light signal to travel from one configuration to the other, but since the configurations and their physical evolutions produce identical experiences, no information need transfer between the two.[5]

Next, imagine that at some point between time $t_2$ and $t_3$, Configuration A is destroyed (or, at the very least, changed to a physical configuration that does not correspond to a conscious state of you). It should be clear that your stream of consciousness, which emerges independently from the physical evolution of Configuration B, is unaffected. Given the symmetry of the hypothetical, destroying Configuration B instead would likewise have no effect on your stream of consciousness.

---

[5] Tappenden (2011) discusses the notion of physically identical people ("doppelgangers") that are spacelike separated: "How can two doppelgangers zillions of lightyears apart whose simultaneity we know, from Special Relativity, is entirely relative to an inertial frame, how can they share a single mind?" His answer? No need for causal connection.

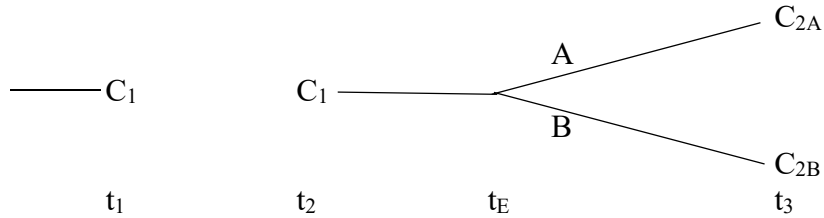

**Fig. 1. Assessing stream of consciousness when paths diverge.**

Now, with reference to Fig. 1, imagine that instead of either configuration being destroyed, an event (such as a quantum measurement event) occurs at some time $t_E$ between $t_2$ and $t_3$ such that Configuration A physically evolves to state $S_{2A}$ at time $t_3$, from which emerges conscious state $C_{2A}$, and Configuration B physically evolves to different state $S_{2B}$ at time $t_3$, from which emerges different conscious state $C_{2B}$. What now? There seem to be three possibilities:

a) Your stream of consciousness ends at $t_E$.
b) Your stream of consciousness follows both paths. At time $t_3$, you experience *both* conscious states $C_{2A}$ and $C_{2B}$.
c) Your stream of consciousness follows a single path, either Configuration A or B. At time $t_3$, you experience either conscious state $C_{2A}$ or conscious state $C_{2B}$ but not both.

Regarding statement a), why would your conscious experience from *both* configurations be extinguished simply because the physical evolution of Configuration B diverged from A? To make matters worse, what if state $S_{2A} = S_2$, which we already know from the previous hypothetical *does* correspond to a conscious state of you (namely, $C_2$)? How could the divergence of a physical system (Configuration B) very far away make any difference to a stream of consciousness that we already know *would* emerge from the physical evolution of Configuration A by itself? Sure, maybe your conscious experience corresponding to the evolution of Configuration B gets cut off, but why *both* of them? And, in any event, if divergence somehow *does* cause your stream of consciousness to end in both configurations, then signals must travel between them. Given that they are spacelike separated, your stream of consciousness certainly could *not* end at $t_E$ as posited and in fact could be delayed arbitrarily long depending on where in spacetime the two configurations were created. Statement a) is false.

Regarding statement b), what would it be like to experience, simultaneously, two *different* streams of consciousness? Just because you may have never before had such an experience and have trouble conceiving of it does not prove that it's false. Perhaps there is a higher conscious plane at which multiple streams of consciousness are possible. But the real problem with b) is Special Relativity. To experience a conscious state emerging from a physical configuration depends *on* the configuration – that is, the information contained in the *physical* state must determine, at least in part, the emergent *conscious* state. Therefore it would be impossible to experience the conscious states corresponding to two spacelike separated configurations of matter. Only statement c) remains, which is consistent with SSCT. A more rigorous proof follows:

I) Assume Physical Consciousness is true.
II) Assume Special Relativity is true.
III) Assume SSCT is false.
IV) By I): Given a conscious entity $CE_1$ whose experience of conscious state $C_1$ depends entirely on physical state $S_1$ of matter $M_1$ and whose stream of consciousness $SOC_1$ depends entirely on a physical evolution of matter $M_1$, it is possible for matter $M_2$, spacelike separated from matter $M_1$, to be configured as state $S_1$ and to evolve differently than matter $M_1$, such that a conscious entity $CE_2$ whose experience of consciousness depends entirely on state $S_1$ of matter $M_2$ experiences a stream of consciousness $SOC_2$, different than $SOC_1$, from a physical evolution of matter $M_2$.
V) By I) and III): the conscious entity $CE_1$ is the same conscious entity $CE_2$ and experiences streams of consciousness $SOC_1$ and $SOC_2$.
VI) By I), the experience by conscious entity $CE_1$ of stream of consciousness $SOC_1$ depends on information associated with the physical evolution of matter $M_1$ and its experience of stream of consciousness $SOC_2$ depends on information associated with the physical evolution of matter $M_2$.
VII) By II), matter $M_1$ and matter $M_2$ are timelike. Contradiction with IV); the three assumptions are incompatible. Therefore, Physical Consciousness and Special Relativity imply SSCT.

The contradiction can be more easily understood by this simple example. Imagine two spacelike-separated physically identical conscious entities whose streams of consciousness begin to diverge. For example, one entity tastes a red apple and thinks, "Delicious!" while the other tastes a green apple and thinks, "Too sour!" It is not possible for any entity to have *both* of these competing experiences because they depend on physical configurations that cannot be connected by any signal traveling at the speed of light. Therefore, if there *is* a stream of consciousness experienced from the initial conscious state, there can only be one. In other words, conscious experience is limited by Special Relativity and must in some sense be localized relative to whatever physical matter created it, and this fact implies SSCT.[6]

One objection to the above hypotheticals is their technological impracticality and/or physical impossibility. It may not be possible, for example, to measure the precise quantum state of a brain (or whatever physical system creates a conscious state), much less to recreate it, in no small part because of quantum no-cloning. But is such precision necessary? If a computer is ever to become conscious, certainly no one doubts that identical software could be run on a functionally equivalent, but physically very different, processor to generate the same consciousness. Analogously, in the case of Physical Consciousness, how good need a copy be so that the same conscious experience emerges?

---

[6] A related argument in which state $S_1$ produced by matter $M_1$ and state $S_1$ produced by matter $M_2$ are assumed to be *timelike* separated, so that one occurs *before* the other, results in a different problem. Because the same conscious entity, if SSCT is false, would experience both streams of consciousness "simultaneously" (i.e., relative to the entity's subjective experience), then one stream of consciousness would depend on information associated with a physical evolution that occurs in the future, implying backward causation.

It may be the case that a large number of variations of a particular physical configuration would produce the same conscious state; i.e., $S_1=\{S_{1a}, S_{1b}, \text{etc.}\}$ is a large set of physical states of matter that produce conscious state $C_1$. All we need for the above hypotheticals is a *sufficiently adequate* copy of a physical state – that is, some member of set $S_1$ – to create the same conscious state $C_1$. Even if that turns out to be exceptionally difficult as a practical matter, we can be assured that eventually, by purely random chance, an adequate copy of $S_1$ should appear in the universe in the form of a so-called Boltzmann Brain. And if that turns out to be physically impossible – if the level of precision needed to duplicate a conscious state exceeded what was physically possible – then SSCT would logically follow anyway.

## 3.     The Multiverse Implies SSCT

Given that Special Relativity is now generally accepted throughout the scientific community, further proofs of SSCT are unnecessary. However, it is interesting to note that SSCT is also a direct consequence of Multiverse theory. Because the universe appears to be exceptionally finely tuned for the existence of life, the desire to avoid appealing to essentially infinitesimal probabilities or divine intervention has led many to embrace the Multiverse theory, which simply suggests in at least one form that every possible physical configuration of a universe that can exist does exist. In other words, instead of having to explain why the gravitational constant or the ratio of the masses of the proton to the electron or the values of a dozen other fundamental constants of nature are just right for the creation of a universe where intelligent life can evolve, it is easier for some to posit the existence of many (perhaps infinitely many) other universes having different constant values.

If true, then there are many identical versions of you in other universes. The approximate distance to the nearest physically identical you has been calculated at around 10 to the power of $10^{29}$ meters and the distance to the nearest entire observable universe identical to our own at around 10 to the power of $10^{120}$ meters. (See, e.g., Davies 2006, p. 178.) These are mind-bogglingly and incomprehensibly vast distances, particularly when compared to the distance to the edge of our own observable universe, a paltry 10 to the power of 26 meters. Of course, given such vast distances, there is no way to observe or communicate with another identical you. While many scientists regard the Multiverse theory neither credible nor scientific due to issues of falsifiability, it is worth showing that it too implies SSCT.

If the Multiverse is true, then there are lots of (perfectly) physically identical versions of me. Because my identity and current conscious experience is a direct consequence of physical state, all of the physically identical versions of me, then, *must* be me, and I must experience all of them. That's not a problem... until some of those physical states begin to diverge. At that point my identities either remain with them, in which case I experience multiple different steams of consciousness, or they don't, in which case I experience just one. But since that must be happening *right now*, the fact that I *don't* experience multiple streams of consciousness implies SSCT. A more rigorous proof follows:

I)      Assume Physical Consciousness is true.
II)     Assume Multiverse theory is true.
III)    Assume SSCT is false.
IV)    By I) and II): for any conscious entity whose experience of consciousness depends entirely on an appropriate configuration of matter, multiple such configurations of matter exist.
V)     By II) and III), at least some of these multiple such configurations evolve differently such that from these different evolutions, the conscious entity experiences different streams of consciousness.
VI)    I am a conscious entity and I do not experience multiple different streams of consciousness. Contradiction with V); the three assumptions are incompatible. Therefore, Physical Consciousness in a Multiverse implies SSCT.

## 4.    SSCT Implies That Consciousness Cannot Be Algorithmic

From a common sense point of view, SSCT simply asserts that there is nothing it's like to experience different streams of consciousness from the same conscious state. So what? It turns out that SSCT implies several surprising conclusions that contradict various closely-held popular and scientific convictions. For instance, one consequence of SSCT is that Strong AI is false.

If two conscious states could evolve along different streams of consciousness, then SSCT implies they cannot be the same conscious state. But all algorithms, except the most boring ones, *can* flow differently, depending on inputs. So imagine a computer running software that becomes conscious by nature of reaching a certain process in the software's algorithm; it then proceeds to experience a first stream of consciousness. Later, the computer is reset to that earlier process. If Algorithmic Consciousness is true, then not only will the computer again be conscious, but it must be the *same* conscious entity subjectively experiencing the *same* moment of awareness and the *same* sensations. We know that it has already experienced the first stream of consciousness, so how could it experience a second (different) stream of consciousness from that *same* state? What would that feel like? It wouldn't feel like anything – because it's not possible. A more rigorous proof follows:

I)      Assume Algorithmic Consciousness is true.
II)     Assume SSCT is true.
III)    By I): because any computable algorithm can be performed on a general purpose digital computer or Turing machine, it is possible in principle for an executing algorithm to be reset to an earlier process of the algorithm's process flow.

IV) By I) and III), the following is possible: A conscious entity $CE_1$ experiences at time $t_1$ conscious state $C_1$ emerging from process $P_1$ of an algorithm. From time $t_1$ to $t_2$, execution of the algorithm flows to process $P_2$ from which emerges the conscious entity $CE_1$'s experience of conscious state $C_2$, resulting in the conscious entity $CE_1$'s experience of a stream of consciousness SOC from state $C_1$ to $C_2$. At time $t_3$, the execution is reset to process $P_1$, from which emerges a conscious entity $CE_2$'s experience of conscious state $C_1$. From time $t_3$ to $t_4$, execution of the algorithm flows to process $P_2$', from which emerges the conscious entity $CE_2$'s experience of conscious state $C_2$', different from $C_2$, resulting in the conscious entity $CE_2$'s experience of a stream of consciousness SOC', different from SOC, from state $C_1$ to $C_2$'.

V) By I) and IV): conscious entity $CE_1$ at time $t_1$ is the *same* conscious entity in the *same* conscious state as conscious entity $CE_2$ at time $t_3$.

VI) By IV) and V): conscious entity $CE_1$ experiences, from conscious state $C_1$, both stream of consciousness SOC and different stream of consciousness SOC'. Contradiction with II). Therefore, SSCT implies that Algorithmic Consciousness is false.

One might object that actual time matters – after all, in the above proof, SOC' happens *after* SOC has completed. However, it should be noted that *two* different computers, both programmed with the same software, could be running process $P_1$ at the same time $t_3$. What happens if the resulting processes then diverge? SSCT – and common sense – tell us that the diverging processes cannot result in diverging streams of consciousness. After all, what would it be like to be the conscious entity at time $t_3$? Further, if these two different computers are spacelike, then simultaneity is relative and the fact of which computer experiences a stream of consciousness first is observer-dependent. Actual time does not actually matter; two entities experiencing the same conscious state will subjectively observe that state simultaneously.

Therefore, Algorithmic Consciousness is false. Because consciousness is not algorithmic, no computer or artificial intelligence will ever become conscious. Further, if consciousness cannot be created by the execution of software on a computer, then it also cannot be simulated, in which case Nick Bostrom's Simulation Argument (2003) is invalid. That of course does not imply that you aren't living *in* a simulated environment – in fact I think it is quite likely that virtual reality technology will improve until humans (voluntarily, I hope) live their lives in a computer-generated environment – but it does imply that your conscious identity itself is not a simulation.

Another implication is that efforts to upload one's mind to a computer – whether to augment one's mental processing power or to guarantee immortality – will likely be fruitless. That isn't to suggest that information contained in one's memory could not somehow be accessed and stored. Rather, a conscious person cannot hope to ever exist as software running on a computer.

## 5.     SSCT Implies That Consciousness Cannot Be Physically Reset

Setting aside the impossibility of simulating or uploading one's consciousness onto a computer, SSCT also implies limitations to consciousness whether instantiated in a physical brain or not. Because Algorithmic Consciousness supervenes on Physical Consciousness, the latter may be true even if the former is false. A nearly identical proof as the one in Section 4 may be given with regard to Physical Consciousness, although now the corresponding statement III), shown below, is suspect:

III') It is possible in principle to reset a physical configuration $S_2$, from which a conscious entity experiences conscious state $C_2$, to an earlier physical configuration $S_1$, from which the conscious entity experiences an earlier conscious state $C_1$.

Statement III) in Section 4 is true because an identical algorithm may be executed on entirely different physical devices. However, if consciousness depends on a sufficiently precise configuration of matter, it may not be physically possible to reset a physical state corresponding to a particular conscious state to an earlier one, which means that the above statement III') may be false. So instead of positing statement III'), I will designate it an *assumption* called Physical Conscious Reset. Thus, the conclusion of the proof in Section 4 applied, instead, to Physical Consciousness instead of Algorithmic Consciousness is: if SSCT is true, then Physical Consciousness is incompatible with Physical Conscious Reset. Assuming Physical Consciousness is true, we will need to find some physical means to explain why and how a conscious physical configuration can't be reset to an earlier conscious configuration – that is, why Physical Consciousness does not necessitate Physical Conscious Reset.[7]

Setting aside logic, the problem with physical resetting can be shown visually in Fig. 2, in which a conscious entity in conscious state $C_1$ emerging from physical state $S_1$ at time $t_1$ is destroyed; in Situation A, the physical state $S_1$ is recreated (not necessarily from the same matter) at time $t_2 > t_1$, from which emerges an entity's conscious state $C_1$ and a stream of consciousness $SOC_A$ from the physical evolution to state $S_2$, from which emerges the entity's conscious state $C_2$, at time $t_3$, which is then destroyed; in Situation B, the physical state $S_1$ is recreated (not necessarily from the same matter) at $t_4 > t_3$, from which emerges an entity's conscious state $C_1$ and a stream of consciousness $SOC_B$ from the physical evolution to state $S_2'$, from which arises the entity's conscious state $C_2'$, at time $t_5$, which is not necessarily destroyed.

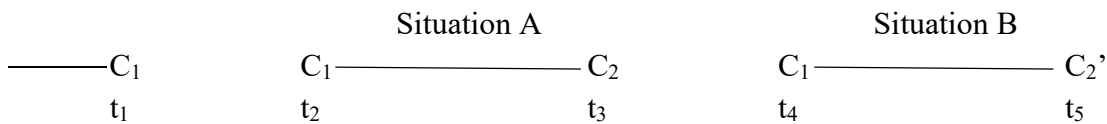

**Fig. 2. Assessing the streams of consciousness in various situations.**

---

[7] Aaronson's "freebit" notion is a possible solution (2016). There are others, but space prevents elaboration.

Without Situation B, there seems to be no logistical problem: the entity's stream of consciousness $SOC_A$ flows continuously from state $C_1$ at $t_1$ to state $C_2$ at $t_3$ via $C_1$ at $t_2$. And without Situation A, the entity's stream of consciousness $SOC_B$ flows continuously from state $C_1$ at $t_1$ to state $C_2$' at $t_5$ via $C_1$ at $t_4$. But when *both* situations occur, we need to figure out what the conscious entity actually experiences as its consciousness flows from time $t_1$.

It cannot actually experience both streams, of course, because SSCT, which is time-independent, prohibits the conscious entity from experiencing more than one stream from the same conscious state ($C_1$). If it could, it would subjectively experience them *simultaneously*, which is not only hard to imagine, but also means the entity would be incapable of experiencing $SOC_A$ until time $t_4$, which could be arbitrarily far in the future. Further, without backward causation or signaling, how could it even be known at time $t_2$ that a Situation B would ever happen?

So perhaps the conscious entity could experience *either* $SOC_A$ *or* $SOC_B$ but not both – if so, how is the selection made? What does the entity actually experience at time $t_1$? Does its stream of consciousness flow toward $C_2$ or $C_2$'? Assume for the moment that the choice is based on chronology: the entity experiences the chronologically *earlier* stream, $SOC_A$, and therefore not $SOC_B$. The problem is that at time $t_4$, physical state $S_1$ is recreated. And since the conscious entity experienced conscious state $C_1$ that emerged from physical state $S_1$ at time $t_2$, then Physical Consciousness guarantees that the *same* conscious entity will experience the *same* conscious state $C_1$ at time $t_4$. Because that physical state $S_1$ then evolves to state $S_2$', defining conscious state $C_2$', the conscious entity will indeed experience $SOC_B$, a contradiction.

But if we assume instead that the entity experiences the chronologically *later* stream, $SOC_B$, and therefore not $SOC_A$, a similar problem arises. The stream $SOC_B$ begins with conscious state $C_1$ that emerges from physical state $S_1$ at time $t_4$, but $SOC_A$ is stipulated to have started at the same state; Physical Consciousness requires that the *same* entity experiences both. An additional problem arises if the entity experiences the chronologically later stream: whether or not the entity experiences a stream of consciousness from $t_2$ to $t_3$ depends on an event at some arbitrary future time $t_4$, resulting in a reverse time dependence.

As a final possibility, perhaps the conscious entity *can* experience both $SOC_A$ and $SOC_B$ because they don't actually chronologically overlap and my earlier assertion – that the entity would subjectively experience them simultaneously – was incorrect. In other words, the conscious entity first experiences stream of consciousness $SOC_A$ and later experiences $SOC_B$. The reset that occurs between times $t_3$ and $t_4$ is effectively a memory erasure such that the conscious entity can proceed to experience stream of consciousness $SOC_B$ without simultaneously (from the entity's subjective perspective) experiencing $SOC_A$. While the entity may *report* after time $t_4$ that it experienced a continuous stream of consciousness directly from $t_1$ to $t_4$ without any intervening experience, it is just deluded, because at time $t_1$ it actually experienced a stream of consciousness that flowed directly from $t_1$ to $t_2$.

The problem here is that $SOC_A$ and $SOC_B$ in fact *do* chronologically overlap, precisely at time $t_1$: stream of consciousness $SOC_A$ flows from $t_1$ to $t_3$ *and* stream of consciousness $SOC_B$ flows from $t_1$ to $t_5$ – i.e., they both flow from time $t_1$. Because physical state $S_1$ at $t_2$ is precisely the same as state $S_1$ at $t_4$ (from which emerge precisely the same conscious state $C_1$ of precisely the same conscious entity), then state $C_1$ at $t_2$ has no better a claim to defining the "real" path of the entity's stream of consciousness than state $C_1$ at $t_4$. When the entity reports after time $t_4$ that it experienced a continuous stream of consciousness directly from $t_1$ to $t_4$, it is correct, because the experience of its stream of consciousness flowing from $C_1$ at time $t_1$ through $C_1$ at time $t_2$ must – *physically must* – be exactly the same experience as its stream of consciousness flowing from $C_1$ at time $t_1$ through $C_1$ at time $t_4$. The apparently seamless transitions from $C_1$ at $t_1$ to *both* $C_1$ at $t_2$ *and* $C_1$ at $t_4$ are in fact real – or, at least, neither is more real than the other.[8]

In other words, through all this analysis, we still don't know what the entity experiences at time $t_1$; does it experience $SOC_A$ or $SOC_B$? If consciousness can be physically reset, then there must be something it's like to be the entity at time $t_1$. Unfortunately, every logical possibility leads to a contradiction, and if there is nothing it's like to be in conscious state $C_1$ at time $t_1$ in Fig. 2, then it is because the stipulated scenario is impossible.

The above arguments show that if SSCT is true, then Situation A and Situation B cannot both occur, and that leads to larger implications. For example, imagine if, in Fig. 2, time $t_1=t_2$ so that physical state $S_1$ is just some arbitrarily chosen state defining an entity's conscious state $C_1$. Situation B is therefore not possible, whether or not $t_4$ comes after $t_3$. In other words: a *copy* of a physical state corresponding to a past conscious experience cannot be created! The Single Stream of Consciousness Theorem does not merely imply that a conscious entity can't be reset to an earlier conscious state, but that an experienced conscious state can't be physically recreated at all.

If I am correct, the good news is that philosophical issues of identity and self-location simply disappear. For instance, if the Philosophy Defense Force tells Dr. Evil that they have created an exact replica of him, which they will torture until he surrenders, Dr. Evil can ignore them, safe in the knowledge that they are liars and that he, and only he, is Dr. Evil. (See Elga 2004.) Also, Roger Penrose can stop worrying about the "duplication problem," whereby a teleportation machine is used to create an exact physical copy of a space traveler on a distant planet but "the original copy of the traveler [is] not destroyed... Would his 'awareness' be in two places at once?" (1989, p. 27)

The bad news is that we then have to find some physical mechanism that prevents certain physical copies from being made. A Boltzmann Brain just isn't physically possible – but why not? Further, what is the physical mechanism that prevents copying or repeating a conscious state? If $C=\{C_1, C_2, etc.\}$ is the set of all conscious states that have already been experienced, and $S_1=\{S_{1a}, S_{1b}, etc.\}$ is the set of all physical states of matter that produce conscious state $C_1$, $S_2=\{S_{2a}, S_{2b}, etc.\}$ is the set of all physical states of matter that produce conscious state $C_2$, and so forth, then the universe must physically contain this potentially massive amount of information and it must be instantly accessible to every physical configuration so as to be useful!

---

[8] The problem becomes even more acute (and interesting) when the recreation event at time $t_4$ begins some time between $t_2$ and $t_3$; however, space prevents elaboration.

# 6. Conclusion

In this paper, I attempted to prove the Single Stream of Consciousness Theorem – that a conscious entity cannot experience more than one stream of consciousness from a given conscious state – by showing that it follows independently from both Special Relativity and the Multiverse theory. I discussed some of the theorem's surprising implications, among them that Strong AI is false, consciousness cannot be algorithmic, and a conscious state cannot be reset to an earlier conscious state nor can it be duplicated.

The implications of SSCT may be counterintuitive and may motivate some to reject it outright. On one hand, it seems intuitively and even empirically true: there is significantly more direct evidence for SSCT – specifically, one's own subjective conscious experiences – than that, for example, consciousness is algorithmic. On the other hand, its implications, only a handful of which were broached in this paper, range from odd to downright incredible. But given that SSCT logically follows from our current understanding of the physical world, it seems that either acceptance *or* rejection presents a quagmire. That should not be shocking, given that so little is known about consciousness and that, when acknowledged, it tends to be an annoying thorn in the side of physics. While I believe the arguments in this paper are sound, and that there is something unique about consciousness that has thus far been inadequately addressed or explained, I am also quite open to the idea that one or more fundamental flaws infect these arguments.

## Appendix: Identity and the Many Worlds Objection to SSCT

Neuroscientist Max Tegmark creates a "quantum gun" designed so that each time its trigger is pulled, it measures the z-spin of a particle in the state $(|\uparrow> + |\downarrow>) / \sqrt{2}$. It then fires a bullet only if the spin is measured down, which would be expected to happen with a likelihood of ½, and otherwise simply makes an audible click. He stands in front of the gun, in what he dubs a "quantum suicide" experiment (1998), and asks his assistant to pull the trigger, after which the wave state of the combined system becomes:

$$(1) \quad \frac{1}{\sqrt{2}}(|\uparrow> \otimes |Tegmark>_{alive} + |\downarrow> \otimes |Tegmark>_{dead})$$

Because experiments are observed to actually have results, some interpretations of quantum mechanics require a nonlinear and irreversible reduction of the system's wave state to a single outcome, such as $|\uparrow> \otimes |Tegmark>_{alive}$ in the present example. However, these "collapse" interpretations are infested by questions of when and how such a reduction occurs. One interpretation that avoids the issue of nonlinear collapse is the Many Worlds Interpretation ("MWI") of Hugh Everett (see, e.g., Wallace 2012) which asserts, simply, that collapse does not occur and every term in a wave state corresponds to an actual world.[9] If true, then the wave state continues to evolve linearly with two terms, one in which the gun did not fire and Tegmark is alive and another where the opposite is true. Tegmark (1998) claims that one implication of this assertion, and even an empirical test of MWI, is a form of immortality: "Since there is exactly one observer having perceptions both before and after the trigger event... the MWI prediction is that [the observer] will hear 'click' with 100% certainty," even though *other* observers can expect to hear "click" with only 50% likelihood. The experiment can be performed as many times as is necessary to prove MWI with arbitrarily high confidence. For instance, the likelihood of surviving 100 experiments if a collapse interpretation of quantum mechanics is true is one in $2^{100}$, while the likelihood if MWI is true is 100%. He further notes that because in "almost all terms in the final superposition" the assistant believes that the observer is dead, if you repeatedly attempt quantum suicide, "you *will* experimentally convince yourself that MWI is correct, but you can never convince anyone else."

This is incorrect. According to MWI, every branch is equally real. So while MWI implies that nearly all branches after a series of quantum suicide events contain people who have witnessed the death of the experimenter, the single world in which the experimenter lives happens to be equally real and therefore contains people who are utterly astounded by his incredible luck. Indeed, the experimenter's assistant, spouse, family, friends, and academic colleagues seem authentic to him – and according to MWI they *are* authentic – differing from other branches only in their incredulity over living in a world in which the experimenter defied such incredible odds.[10]

---

[9] These orthogonal terms are often characterized as different "branches" of the universe, or even entirely different universes. The so-called Splitting Worlds View by DeWitt (1970) is a popular interpretation of MWI, although Tegmark points out that this misrepresents MWI in part because the terms could interfere over time. Nevertheless, what are we to make, conceptually, of Expression (1) if not different branches or even universes?

More importantly, the problem of *identity* infects both the quantum suicide thought experiment as well as MWI[11] in general. Tegmark (1998) implies that because there is one conscious observer before the experiment and one conscious observer after the event, it must be the *same* observer. But imagine that on quantum suicide event #58, for example, the experimenter sneezes, so that any bullet shot from the gun would graze his head and cause pain instead of death. Which of the two observers – the unharmed one or the injured one – could the experimenter expect to be? After event #58, there are now two branches and, presumably, two conscious observers who can each equally claim to be the pre-event experimenter. But this does not answer the question of which of the two observers *he became*. This is a massive problem with MWI, so I'll elaborate.

Assume that an experimenter, $Alice_1$, performs a quantum mechanical spin measurement on a particle, after which emerges experimenters $Alice_2\uparrow$ (who observes an "up" measurement) and $Alice_2\downarrow$ (who observes a "down" measurement). Is $Alice_1$ the same person as $Alice_2\uparrow$? As $Alice_2\downarrow$? As both? As neither? What stream of consciousness does $Alice_1$ experience as she does the experiment? Which result will $Alice_1$ see? Which of the two post-measurement experimenters should $Alice_1$ *expect* to become?

MWI contradicts SSCT because if all terms in a quantum wave state correspond to an actual world, then $Alice_2\uparrow$ and $Alice_2\downarrow$ must be equally real. Given that a system's wave state is believed to be its *complete* physical description, if MWI is true, then $Alice_1$ cannot expect her identity to flow into just one or the other; if she could, then a nonphysical element (which would be lethal to MWI) must be introduced because nothing in the quantum wave state would determine one path over the other. To avoid the potential for dualism, Peter Lewis (2007) argues that $Alice_1$ can expect to become *both* $Alice_2\uparrow$ *and* $Alice_2\downarrow$ in separate branches, but later denies the concept of identity altogether: "I am not convinced that the pronoun 'I' picks out a *person* in any deep metaphysical sense...". However, each assertion, while consistent with MWI, is intellectually jarring and requires further elaboration. First, what would it mean to *become* two different people with different minds and experiences? What would that feel like? If MWI is correct and if personal identity flows to *all* branches, then it has already happened to me uncountably many times; why have I not noticed? Why do I observe a continuous and singular stream of consciousness? Second, how can the word "I" not identify someone? As I stare out from behind my eyes, as I introspect, it is clear to me that the word "I" refers to something very specific: my identity, my awareness, my consciousness.

---

[10] David Lewis (2004) makes the same mistake. On branching, he states, "All your future selves, on all your branches, are equally real and equally yours." On quantum suicide: "Your evidence against collapse, if you gain it, ... [cannot be shared] with a bystandard."

[11] Because DeWitt's interpretation of MWI requires infinite mass and energy to support continuously splitting worlds, a clear violation of conservation laws, several "many minds" interpretations ("MMI") have been proposed. Perhaps the best known, the Many Minds View ("MMV") of Albert and Loewer (1988) proposed that brain states that are associated with mental states are associated with an infinite set of minds so that at each measurement, every possible outcome is observed by one or more minds. However, because "supervenience [of the mental on the physical] fails in relation to the question which minds end up tracking which terms of the state vector as new superpositions arise" (Lockwood 1996), MMV maintains dualist commitments, not satisfying to those who seek purely physical explanations.

Lockwood (1996) agrees that identity flows into all branches and states that the remarkable conclusion of being "*literally* in two minds" is "no *more* remarkable... than the already utterly mysterious fact that, at a given time, there is even *one* 'what is it like to be' associated with my brain." I find his assertion not credible, given that he has, presumably, amassed a lifetime's worth of evidence that he has one mind and not a shred to suggest that he occupies two or more; it would certainly be more remarkable to discover that the latter was true. Further, he does not comment on the "what it's like" identity issue *prior* to the experiment. David Papineau (2004) addresses this deficiency: if MWI is correct and he opens Schrodinger's box, "I have no uncertainty about the impersonal structure of the future. I know for sure that it will contain a successor of David Papineau who sees a live cat, and one who sees a dead cat. All I am unsure about is what *I* will see."

One way to avoid MWI's identity crisis is to simply deny Tegmark's assertion that "there is exactly one observer having perceptions... before... the trigger event." Saunders *et al.* (2010) suggest that every possible branch already exists; prior to a branching event, multiple versions of the experimenter already exist, but because they are identical, they cannot self-identify. Observer Alice, for example, remains with her own branch before and after the branching event, but until the branching event, she cannot "know which of these branching persons is she." No dualistic theory is required to explain why her stream of consciousness follows one branch and not another, nor need we explain how a single mind fissions into two. The problem with the Saunders *et al.* solution is that the observers prior to the branching event are *physically identical*, which means that there is no sense in which they are different or have their own individual identities. $Alice_A$ and $Alice_B$, prior to branching, must be the *same person*; if not, then nonphysical facts must be introduced to distinguish them. It is not that $Alice_A$ just cannot yet *tell* that she's not $Alice_B$; rather, because they are physically identical, there is *no sense* in which $Alice_A$ is NOT $Alice_B$. $Alice_A$ inevitably experiences what it's like to be $Alice_B$ and vice versa; every subjective experience of $Alice_A$ is experienced by $Alice_B$; there are not two observers at all; there is only one person and one identity. We are left with the same unanswered question at branching: which person does Alice *become*?

Finally, Ismael (2003) addresses the identity problem of MWI head-on: just before a measurement, "you are wondering what sort of result *you* will experience ... (not any of your externally indistinguishable counterparts, but *you*)...". Unfortunately, she explicitly ignores her own question ("perhaps I needn't be able to say beforehand which one of the post-measurement observers is *me*") and instead focuses on making sense of the Born probabilities: "they tell her how surprised she should be...". According to Ismael, when $Alice_1$ branches into $Alice_2\uparrow$ and $Alice_2\downarrow$, there is no identity, no stream of consciousness, that flows from $Alice_1$. Rather, analyzing the situation *post*-measurement, $Alice_2\uparrow$ and $Alice_2\downarrow$ can each equally claim to *be* $Alice_1$.

Ismael's exclusively backward-looking approach fails to solve the problem. Assume that I am $Alice_1$ and about to perform a quantum mechanical spin measurement on a particle. The post-measurement versions, $Alice_2\uparrow$ and $Alice_2\downarrow$, may both exist or only one may exist, but whoever does exist after the measurement will have a specific temporal relationship to me, will share my memories, and will identify as me. It is reasonable to ask, prior to the measurement, which of the two I will become, and it is *also* reasonable to wonder what I will experience at the moment of measurement. If I do have a stream of consciousness and it chooses one path, then I will experience a continuous stream of consciousness from $Alice_1$ to $Alice_2\uparrow$ (or $Alice_2\downarrow$). Of course, I cannot be certain before doing the experiment that my consciousness will choose one path, but for that matter I cannot be certain that there actually is any such thing as a stream of consciousness. Still, given the evidence of my experience and memories, it is reasonable to believe that I will do the measurement and then observe either up or down. In a continuous flow, I in fact proceed to do the measurement and then observe, say, up – or at least that is what I *think* happened, reporting as $Alice_2\uparrow$. But the flowing of identity from $Alice_1$ to $Alice_2\uparrow$ conflicts with MWI, so what really happened? MWI apologists will, after the fact, explain my experience as such: now that the measurement is past, there is an $Alice_2\uparrow$ and $Alice_2\downarrow$, both of whom are conscious, self-aware, and identify as Alice, and both of whom would report the memory of a seamless flow of consciousness from $Alice_1$.

There are two fatal problems with this story. The first is that while we have an account of what each of $Alice_2\uparrow$ and $Alice_2\downarrow$ recalls experiencing *after the fact*, we have no account of what $Alice_1$ *actually experienced*. We have no idea what it's like to be $Alice_1$ at the moment of branching. However, if we take MWI at face value, then $Alice_2\uparrow$ specifically remembers the experience of being $Alice_1$ at the moment that she became $Alice_2\uparrow$ (an experience I'll denote "$1\rightarrow2\uparrow$"), and $Alice_2\downarrow$ specifically remembers the experience of being $Alice_1$ at the moment that she became $Alice_2\downarrow$ (an experience I'll denote "$1\rightarrow2\downarrow$"). $Alice_1$, who was *one* person, cannot have perceived, simultaneously, two mutually exclusive experiences. Further, she could not have perceived just *one* of those experiences, as that would require the sort of nonphysical reduction information that MWI was designed to reject. After all, if $Alice_1$ actually experienced, say, $1\rightarrow2\uparrow$ at the moment of measurement, then we have our answer as to which of $Alice_2\uparrow$ and $Alice_2\downarrow$ inherited $Alice_1$'s identity! Therefore, $Alice_1$ could not have experienced *either* $1\rightarrow2\uparrow$ *or* $1\rightarrow2\downarrow$.

But that's just silly. Imagine that $Alice_0$ is about to perform a measurement on a *deterministic* system for which only one possible post-measurement version of the experimenter ($Alice_1$) would exist. In that case, there is no reason to believe that $Alice_0$ would not experience $0\rightarrow1$. In other words, $Alice_0$ would experience $0\rightarrow1$, $Alice_0$'s identity would flow into $Alice_1$, and $Alice_1$ would specifically remember $0\rightarrow1$. So how could it be that the measurement of a *non*-deterministic system has the effect that $Alice_1$ does not, in fact, experience what *any* future versions of her remember experiencing? Why is $Alice_1$'s memory reliable but $Alice_2\uparrow$'s memory *guaranteed* to be wrong? MWI implies a weird type of consciousness in which one's memories are not merely untrustworthy but are *certain* to be wrong.

The second problem with the MWI explanation is this: *I, in fact, observed a measurement of up.* The MWI response: "Yes, of course, but that's because you're Alice$_2\uparrow$. Alice$_2\downarrow$ observed a measurement of down." That may or may not be true, but it does not answer the underlying question: *why* am I Alice$_2\uparrow$? Why am I aware of the thoughts of Alice$_2\uparrow$? Why am I seeing out of the eyes, and introspecting the feelings, and experiencing the awareness, of Alice$_2\uparrow$ and not Alice$_2\downarrow$? *Why* am I Alice$_2\uparrow$ and *not* Alice$_2\downarrow$?

At this point MWI apologists may simply aver that I don't understand MWI. "You are Alice$_2\uparrow$," they repeat, annoyed, "and someone else is Alice$_2\downarrow$, and that explains why you see out of Alice$_2\uparrow$'s eyes. End of story." And to that I would reply that they don't understand *identity*. "To you," I would say, "Alice$_2\uparrow$ and Alice$_2\downarrow$ are essentially identical, and except for the serendipity of sharing the same world with Alice$_2\uparrow$, you don't see anything special about Alice$_2\uparrow$. But to *me*, Alice$_2\uparrow$ and Alice$_2\downarrow$ are *extremely* different. Specifically, Alice$_2\uparrow$'s eyes are giving me visual information, while Alice$_2\downarrow$'s are not. I can read Alice$_2\uparrow$'s thoughts and feel Alice$_2\uparrow$'s feelings, but not those of Alice$_2\downarrow$. There is something very different about Alice$_2\uparrow$, and even though you cannot know it, I do. MWI provides no plausible physical explanation for why I experience the feelings of Alice$_2\uparrow$ and not Alice$_2\downarrow$ – for why I *identify* as Alice$_2\uparrow$ and not Alice$_2\downarrow$." It seems to me that there are four possible responses to the MWI identity problem, at least one of which must be true:

a) There is no "I". It is a false illusion that I am Alice$_2\uparrow$ and not Alice$_2\downarrow$.
b) There is an "I", but I am in fact both Alice$_2\uparrow$ *and* Alice$_2\downarrow$.
c) Consciousness is not purely physical. Something beyond physical facts determined that I would be Alice$_2\uparrow$ and not Alice$_2\downarrow$.
d) MWI is false.

Statement a) is, in my opinion, simply unfathomable. While zombies may engage in sophisticated philosophical debate over whether identity is an illusion, none of them can experience the overwhelming subjective evidence that my identity is not an illusion.

Statement b) simply does not fit the empirical (if subjective) evidence: the fact that I see out of the eyes and experience the awareness of Alice$_2\uparrow$ and not Alice$_2\downarrow$. Clearly there is *something* different between Alice$_2\uparrow$ and Alice$_2\downarrow$. Even if Alice$_2\downarrow$ exists and is conscious, the "I" that experiences Alice$_2\downarrow$ is not me. I experience Alice$_2\uparrow$ and I do *not* experience Alice$_2\downarrow$, so statement b) is false.

Left only with statements c) and d), if MWI is correct, then something had to distinguish *my* identity upon the branching of Alice$_1$. The problem here is that MWI requires that the quantum wave state is the *complete* description of the universe, but nothing in the wave state indicates or *could* indicate which of Alice$_2\uparrow$ or Alice$_2\downarrow$ will inherit my identity. In other words, statement c) implies statement d), thus disarming MWI as a credible threat to SSCT.[12]

---

[12] MWI also suffers from an evidentiary problem. Because MWI implies that conscious Alice$_1$ branches into conscious Alice$_2\uparrow$ and conscious Alice$_2\downarrow$, it is instructive that I have excellent evidence that Alice$_2\uparrow$ exists and is conscious (because I am her) but *no* evidence that Alice$_2\downarrow$ even exists, much less that she is conscious. Significantly, as MWI

is postulated to not allow communication between branches, the evidence for Alice$_2$↓'s consciousness not only *does not* exist but *cannot* exist.  The inherent unfalsifiability built into the very fabric of MWI should give scientists pause.

# References

Aaronson, S., 2016. The Ghost in the Quantum Turing Machine. In: *The Once and Future Turing: Computing the World*. Cambridge University Press.

Albert, D. and Loewer, B., 1988. Interpreting the many worlds interpretation. *Synthese*, *77*(2), pp.195-213.

Bostrom, N., 2003. Are we living in a computer simulation?. *The Philosophical Quarterly*, *53*(211), pp.243-255.

Davies, P., 2006. *The Goldilocks Enigma: Why Is the Universe Just Right for Life?*. Houghton Mifflin.

DeWitt, B.S., 1970. Quantum mechanics and reality. *Physics today*, *23*(9), pp.30-35.

Elga, A., 2004. Defeating Dr. Evil with self-locating belief. *Philosophy and Phenomenological Research*, *69*(2), pp.383-396.

Ismael, J., 2003. How to combine chance and determinism: Thinking about the future in an Everett universe. *Philosophy of Science*, *70*(4), pp.776-790.

Lewis, D., 2004. How many lives has Schrödinger's cat?. *Australasian Journal of Philosophy*, *82*(1), pp.3-22.

Lewis, P.J., 2007. Uncertainty and probability for branching selves. *Studies In History and Philosophy of Science Part B: Studies In History and Philosophy of Modern Physics*, *38*(1), pp.1-14.

Lockwood, M., 1996. 'Many Minds'. Interpretations of Quantum Mechanics. *The British Journal for the Philosophy of Science*, *47*(2), pp.159-188.

Papineau, D., 2004. David Lewis and Schrödinger's cat. *Australasian Journal of Philosophy*, *82*(1), pp.153-169.

Penrose, R., 1989. *The Emperor's New Mind: Concerning Computers, Minds, and the Law of Physics*. Oxford University Press.

Saunders, S., Barrett, J., Kent, A. and Wallace, D., 2010. *Many Worlds?: Everett, Quantum Theory, & Reality*. Oxford University Press.

Searle, J.R., 1980. Minds, brains, and programs. *Behavioral and brain sciences*, *3*(3), pp.417-424.

Tappenden, P., 2011. A metaphysics for semantic internalism. *Metaphysica*, *12*(2), p.125.
Tegmark, M., 1998. The interpretation of quantum mechanics: Many worlds or many words?. *Fortschritte der Physik: Progress of Physics*, *46*(6-8), pp.855-862.

Wallace, D., 2012. *The Emergent Multiverse: Quantum Theory According to the Everett Interpretation*. Oxford University Press.

Zuboff, A., 1990. One self: The logic of experience. *Inquiry*, *33*(1), pp.39-68.